# Spiritual Intelligence's Role in Reducing Technostress through Ethical Work Climates

**Saleh Ghobbeh**, Faculty of Entrepreneurship, University of Tehran, Tehran, Iran. salehghobbeh@ut.ac.ir

**Armita Atrian**, Department of Management, Faculty of Administrative Sciences, Imam Reza International University, Mashhad, Iran. armita.atrian@imamreza.ac.ir

## Abstract

This study explores the impact of spiritual intelligence (SI) on technostress, with a focus on the mediating role of the ethical environment. In an era where technological advancements continually reshape our work and personal lives, understanding the interplay between human intelligence, well-being, and ethics within organizations is increasingly significant. Spiritual intelligence, transcending traditional cognitive and emotional intelligences, emphasizes understanding personal meaning and values. This paper investigates how higher levels of SI enable individuals to integrate technology into their lives without undue stress, and how a robust ethical environment within organizations supports and amplifies these benefits. Through a comprehensive review of literature, empirical research, and detailed analysis, the study highlights the protective role of SI against technostress and the significant influence of an ethical climate in enhancing this effect. The findings offer valuable insights for organizational strategies aimed at promoting a harmonious, stress-free workplace environment.

**Keywords**: Spiritual Intelligence, Technostress, Ethical Environment, Organizational Well-being, Technological Integration.

## Introduction

In the rapidly evolving landscape of modern technology, the interplay of human intelligence, psychological well-being, and ethical practices within organizations has become increasingly significant. This paper delves into the realm of spiritual intelligence (SI), a concept that transcends traditional cognitive and emotional intelligences, emphasizing a deeper understanding of personal meaning and values. The burgeoning field of spiritual intelligence studies suggests its potential to aid individuals in navigating complex and technologically infused environments. In a world where technological advancements continually reshape our work and personal lives, understanding the role of SI in coping with technostress—a byproduct of our ever-increasing reliance on technology—becomes crucial.

Technostress, a term coined in 1984, captures the negative psychological impact stemming from the inability to cope with new computer technologies in a healthy manner. This phenomenon has gained prominence in organizational studies, highlighting the need to address the challenges posed by rapid technological advancements in the workplace. The paper investigates the mitigating role of spiritual intelligence in reducing technostress, positing that individuals equipped with higher levels of SI are better positioned to integrate technology into their lives without undue stress. The study also explores the concept of the ethical environment within organizations, examining how it influences the relationship between SI and technostress.

The ethical environment in organizations, characterized by shared values, ethical leadership, and supportive policies, plays a pivotal role in shaping employees' experiences with technology. This paper posits that a robust ethical environment supports and amplifies the benefits of spiritual intelligence in mitigating technostress. Organizations can reduce the conflict and stress associated with technological demands by fostering a workplace culture where ethical practices are emphasized. This study aims to provide a comprehensive understanding of the interplay between spiritual intelligence, technostress, and the ethical environment, offering valuable insights for organizational strategies aimed at promoting a harmonious, stress-free workplace.

## Literature Review

### Spiritual Intelligence

The concept of spiritual intelligence is a relatively new intellectual development that emerged after Howard Gardner's multiple intelligences, which he introduced in his book "Frames of Mind" in 1983. The evolution of spiritual intelligence can be traced back to the development of emotional intelligence. Emotional intelligence was first introduced in 1985 by Payne and later presented as social intelligence by Salovey and Mayer. Goleman explored and utilized its application in global business in his 1995 book.

Afterward, spiritual intelligence was introduced; however, the exact origin and the person responsible for coining this term remain unclear (Zarei et al., 2011: 73). Gardner defines intelligence as a set of abilities used for problem-solving and creating new products. In general, intelligence enables an individual to adapt to their environment and provides them with coping strategies for problem-solving. Furthermore, the ability to recognize problems, propose potential solutions, and discover efficient problem-solving methods are characteristic of individuals with high intelligence. Scholars have classified intelligence based on Gardner's definition into various categories. According to one classification, human intelligence is divided into three categories: cognitive intelligence or IQ, emotional intelligence or EQ, and spiritual intelligence or SQ. The concept of spiritual intelligence was first introduced in 1996 by Stevens and later expanded upon by Emmons in 1999. Emmons defines spiritual intelligence as the adaptive use of spiritual information to facilitate everyday problem-solving and the process of achieving goals (Gardner, 2000, pp. 27-37).

Wolman (2013) states that spiritual intelligence is the capacity of humans to question the meaning of life while simultaneously experiencing a continuous connection between each of us and the world we live in.

According to Zohr (2012), through spiritual intelligence, humans engage with issues of meaning and value, addressing these problems and shaping their actions and lives within a broader context that is richer and more profound in terms of meaning. Zohr (2012) points out that spiritual intelligence has the ability to access higher-level meanings, values, more sustainable goals, and unconscious aspects of oneself. The capacity to place these meanings, values, and goals in creating a richer and more creative life is inherent in spiritual intelligence. The understanding of meaning, values, and goals created by spiritual intelligence leads to the formation of spiritual capital within an individual. Cisco (2002) identifies a crucial aspect of spiritual intelligence in problem-solving ability. He believes that spiritual intelligence enables individuals to see a larger picture and align their actions with a broader context that leads to the meaning of life. A person can recognize and solve problems related to meaning and value through spiritual intelligence.

Yang elaborates: "Spiritual intelligence is the human capacity to explore and ask ultimate questions about the meaning of life while simultaneously experiencing a seamless connection between each of us and the world we live in. Through spiritual intelligence, we address problems by considering their place, meaning, and value. It is the intelligence through which we imbue our actions with meaning and understanding, allowing us to be aware of which of our actions hold more significance and which path in our lives is higher and superior, enabling us to model and shape our lives accordingly" (Yang, 2007: 1000). Other definitions of spiritual intelligence are presented in the table below.

What is apparent from the initial exploration of the definitions provided for the concept of spiritual intelligence is the emphasis of scholars in this field on the adaptive response of spiritual intelligence in everyday life. A significant portion of this adaptive response is seen in the answers to existential questions and the discovery of meaning and purpose in activities and life events. In fact, having meaning and purpose in life appears to be a common denominator in most of these definitions and seems to be the most essential feature of spiritual intelligence. This feature can also serve as a distinguishing factor of this intelligence from other recognized types of intelligence.

Self-awareness is another factor that has been considered as one of the essential features of spiritual intelligence in the definitions provided by most scholars in this field. Deep self-awareness involves a gradual understanding of various layers and dimensions of oneself. Self-reflection and personal awareness can deepen through introspection, setting individual goals and aspirations, paying attention to messages in dreams and intuitive signs, and finding meaning in external experiences (Cisco, 2008).

Various components of spiritual intelligence, as perceived by different scholars, have been studied and will be discussed further below. Zohr (2012) believes that spiritual intelligence signifies the ability to adapt, choose, and shape one's surroundings. Consequently, having a comprehensive understanding of life, spiritual intelligence assists an individual in recognizing deeper meaning in events and experiences and the connection between them. A person can utilize their spiritual intelligence to shape and reinterpret experiences. This process, in phenomenological terms, is capable of adding deeper personal meaning and value to events and experiences.

The non-specific and holistic nature of spiritual intelligence, along with its use of symbolism, empowers individuals to benefit from broader and deeper perceptions. This aspect enriches an individual's relationships and activities in daily life, making them more meaningful and fulfilling. Furthermore, the movement towards self-realization and spiritual growth is more closely associated with spiritual intelligence than with the need for conformity or societal norms. Consequently, individuals who integrate spiritual intelligence into their daily lives may have unconventional lifestyles or different approaches to life, driven by their inclination to live authentically and cohesively.

According to Zohr (2000), spiritual intelligence can be cultivated through various exercises, emotional regulation, and ethical behavior. He believes that spiritual intelligence grows through primary endeavors such as questioning the essence of situations, posing "whys" to issues, and establishing connections between events. Additionally, learning, recognizing, and listening to guiding intuitions, being contemplative, increasing self-awareness, learning from mistakes, and being truthful to oneself contribute to enhancing spiritual intelligence. From Zohr's perspective (2000), individuals with high spiritual intelligence possess traits such as wisdom, completeness, empathy, holism, accuracy, and flexibility. Spiritual intelligence transcends an individual's physical and cognitive interactions with their environment, delving into the realm of intuition and sublime perspectives. It is believed that spiritual intelligence and emotional intelligence are interdependent, with the growth of spiritual intelligence significantly strengthening emotional intelligence. In fact, these two intelligences reinforce each other positively.

Zohr (2012) delves into the concept of intelligence, highlighting its essence as the ability to adapt, choose, and shape one's surroundings. Within this framework, spiritual intelligence emerges as a profound guide, offering individuals a holistic perspective on life. By embracing this holistic outlook, individuals gain the capacity to perceive a deeper meaning within events and experiences and recognize the intricate connections between them. Spiritual intelligence becomes a transformative tool, empowering individuals to understand and reinterpret their life experiences. This transformative process, rooted in the realm of phenomenology, imbues events and experiences with a profound personal significance, adding layers of depth and value to one's journey.

Building upon these insights, Zohr and Marshall (2004) have outlined nine fundamental characteristics that embody a well-developed spiritual intelligence. Firstly, flexibility emerges as a key trait, signifying the individual's ability to adapt to diverse situations and challenges. A high level of self-awareness is another cornerstone, fostering a deep understanding of one's identity and purpose. Moreover, the capacity to confront and utilize pain reflects resilience, enabling individuals to face adversity and transform it into valuable life lessons. Similarly, the ability to transcend pain elevates one's spiritual intelligence, emphasizing the transformative power of resilience in the face of suffering. According to Bai et al. (2023), This heightened emotional resilience can be helpful in coping with stress and uncertainty, and it may offer solace during trying times.

Additionally, drawing inspiration from dreams and values underscores the importance of intuition and ethical foundations in decision-making and personal growth. The inclination to refrain from harming or offending others speaks to the empathy and compassion embedded in spiritual intelligence. Moreover, the ability to perceive connections between various elements highlights a holistic worldview, enabling individuals to discern patterns and interrelations in the complexity of life. The clear inclination to ask profound questions about existence, encapsulated in the tendency to inquire "why" and seek fundamental answers, reflects a contemplative and introspective nature. Lastly, the independence of the field emphasizes the ability to transcend specific contexts and societal norms, allowing individuals to engage in activities that challenge conventional boundaries.

Collectively, these characteristics serve as the foundation of a well-developed spiritual intelligence. They equip individuals with the wisdom and discernment necessary to navigate life's intricate tapestry, infusing their experiences with purpose and meaning. In essence, spiritual intelligence becomes the guiding light, illuminating the path toward profound self-discovery, empathy, and a harmonious connection with the universe.

**Technostress**

"After the technological revolution, the number of organizations utilizing technology increased gradually, to the point where today, it is rare to find an individual or organization not engaged with technology (Hessari, 2023). Technology has accelerated work processes, improved communication between employees and managers, enhanced employee control and production processes by managers, increased efficiency and productivity, and made tasks more effective, among many other benefits. However, as Bai and Vahedian (2023) mentioned, "Dependency on modern technology has brought challenges for organizations in recent years, with numerous inadvertent organizational consequences." The use of technology brings social costs for organizations, including maintenance expenses and changes in work systems. Consequently, there has been significant attention to this issue in recent decades. Many scientists and researchers have explored the negative impacts of technology, unanimously considering technostress as a significant drawback. The term 'technostress' was coined in 1984 by clinical psychologist Dr. Craig Brod. Technostress is a modern ailment resulting from the inability to cope with new computer technologies in a healthy manner. It is a problem arising from the inability to adapt to or cope with information and communication technology (ICT). Technostress includes five components: 1. Techno-overload, where users are forced to work faster and longer due to information and communication technology. 2. Techno-invasion, where ICT users feel they can be constantly connected to work, arising from the blurring of boundaries between work and personal life. 3. Techno-complexity, where ICT users feel their skills are insufficient to cope with the complexities of information and communication technology, forcing them to spend time and effort to learn and understand various aspects of ICT. 4. Techno-insecurity, where ICT users feel threatened that they may lose their jobs or be replaced by others who are better at ICT. 5. Techno-uncertainty, where ICT users feel uncertain and anxious about the changing nature of information and communication technology.

Technostress, in the realm of modern organizations, presents a significant challenge that needs to be addressed. Technology has rapidly evolved, giving individuals the means to be more efficient both in the workplace and at home. However, this increased efficiency comes at a cost, known as 'technostress.' Technostress, or the inability to adapt to new computer technologies due to dependency on them, impacts individuals and organizations negatively. Brod regarded technostress as a modern ailment resulting from humans' inability to adapt to new computer technologies in a healthy way. Although Brod viewed technostress as a disease, other researchers have considered it as an inability to cope with the changes imposed by technology. This decreased efficiency affects individuals' performance in their tasks and, consequently, the organizational success. Technology, especially information technology, is rapidly changing, and organizations that do not keep up lose their competitive edge. The use of new technologies can create stress for employees and negatively impact productivity. It may be cost-effective for an organization, but accessing new information for timely decision-making is essential. Consequently, managing technostress requires addressing 'change,' not just in technical aspects but also in areas such as machinery, software, and networks. Moreover, technostress has been used to identify an individual's feelings when overwhelmed by technology. Other terms synonymous with technostress used by researchers include technophobia, computer phobia, computer anxiety, and computer stress. The research conducted so far has shown that technostress can affect individuals and organizations in various ways. Research articles can be categorized into the following groups:

These studies primarily focus on how technostress impacts the performance of an organization or business. Technostress significantly and negatively influences both Perceived Organizational Commitment and individual innovation (Hessari et al., 2023). The effects of technostress can be classified into three areas. In the organizational context, most articles delve into information volume and data processing issues. Libraries are significantly impacted by the volume of information that needs to be stored and categorized. Employees can become fatigued and disheartened by technology. Technology fatigue causes employees to lose their efficiency. Managing technostress can be challenging for an organization. Will and Rosen found scientific evidence indicating that technostress can lead to information overload, excessive information, loss of motivation, and job dissatisfaction. In the same domain, another study by Wood (2001) and Hogg (2006) investigates how organizations cope with technostress, proposing methods and solutions to manage technology within organizations to reduce technostress. Identifying and managing technostress enhances organizational efficiency.

In the healthcare domain, according to Popovich's research (1994), if technostress levels are high, access and utilization of technology for an organization will be problematic. Organizational employees must accept and adapt to new technology to work with it. According to this research, another fundamental issue related to technostress is health and psychological aspects. Based on the statements of Tams et al. (2005), technostress can lead to sleep problems and depression. This, in turn, affects many other aspects of life, such as work and family. According to Pransky's research (2002) and colleagues, technostress can directly impact individuals' physical and mental pressures, such as high blood pressure, heart diseases, and skeletal disorders. In the same domain, based on Arntz's research (1997), individuals heavily reliant on new technologies such as computers in their work often experience that if these individuals fail to manage stress properly, this trend can have negative effects on their individual and social activities, leading to problems in the workplace. Therefore, technostress negatively affects employees' physical and mental health, resulting in inefficiency and reduced employee performance. Another major concern for organizations is productivity. By using technology, organizations expect an increase in their productivity levels. However, if technostress increases due to technology, productivity can be affected in various ways. Advocating research in the productivity domain was one of the first studies to demonstrate the relationship between technostress and productivity. This connection serves as a good foundation for organizations to base their work on.

**Ethical environment**

The root of the word "ethics" comes from the Greek word "Ethikos," meaning "authority of customs and traditions" and the individual's cultural beliefs (Wood & Renshaw, 2003, pp. 350-344). In general, the term ethics, especially organizational ethics, deals with standards related to the rightness and wrongness of behaviors (Fisher & Boone, 2007, p. 1561).

Work ethics emphasized by religious doctrines such as Protestantism align with the principles of Islam, emphasizing hard work, dedication, creativity, and avoiding unethical methods and behaviors in accumulating wealth and advancing work based on cooperation and friendship. However, in Islam, emphasis has been placed on legitimate and pure intentions rather than outcomes (Yousef, 2001).

Today, ethics in organizations is discussed under various titles, such as ethical organizations, ethical leadership, organizational ethics, professional ethics, and more. However, one of the newest concepts introduced by Victor and Cullen was ethical work climate, which led to the formation of the ethical work climate model. In fact, the first experimental and theoretical study on ethical work climate was conducted by Victor and Cullen (Fritzsche, 2000).

The attention to the role of ethics and moral values in various aspects of life has always existed throughout human history and will continue to do so. This attention to ethics is because human nature has an undeniable inclination towards divine and human values. However, according to Ali and Al-Kazemi (2007), beliefs about business ethics have varied over time and among different countries. The role of ethics in work and organizational environments has been well-documented through various research studies (Ali & Al-Kazemi, 2007).

**Introduction of Ethical Work Climate:**

The ethical climate is one of the primary factors shaping organizational relationships and employee attitudes. Ethical climate encompasses patterns guiding employee behavior and reflects the ethical characteristics of an organization. It can be seen as the general perception of specific organizational behaviors and practices that have ethical content. Scholars like Shapri and Rosenblatt (2010) have conducted extensive research over the past two decades, with much of it led by scholars like Cullen.

**Ethical work climate** was introduced for the first time in 1987. This construct is defined as the "common perceptions about what is ethically correct behavior and how ethical issues should be handled in the organization" (Victor, 2003). Individuals in organizations have come to understand that the existing norms and organizational rules within the organizational structure and policies create a specific work environment. The ethical climate is a type of work environment that reflects the organization's policies, methods, and ethical consequences. Revealing the relationship between organizational ethical climate and employee attitudes and behaviors is one of the most critical topics in organizational studies. Employees who perceive an ethical climate prevailing in the organization consider their interactions and relationships fair. This perception enhances employee satisfaction and commitment.

In situations where ethical codes are implemented, high levels of ethical behavior emerge in organizations. In other words, when codes of ethics become integral to employees' professional knowledge, their ethical decision-making is significantly influenced. Ethical codes indicate which behaviors are appropriate within a group or for individual members both inside and outside the organization. Ethical codes are norms and values that guide employees' thoughts and actions through internal organizational culture criteria.

Effective and well-understood ethical codes are highly likely to transform into ethical behavior. They also reduce unethical behaviors at the organizational level and help resolve ethical issues (Nadi & Hadaeghi, 2011).

Organizational climate is a broad term that refers to employees' perception of the general work environment in an organization; it is like a protective umbrella nurturing subsets like the ethical climate. Research has shown that a healthy organizational climate profoundly impacts employees positively. Therefore, its constituting factors and elements include honesty (transparency), standards, responsibilities, flexibility, rewards, and group commitment. When the levels of these dimensions are high among employees, individuals are inspired by their work environment, making the workplace enjoyable and profitable. A more positive organizational climate facilitates easier human interactions. In contrast, closed, fearful, and negative climates breed distrust, fear, distance, and resentment among individuals. Thus, a positive and open atmosphere ensures the mental well-being of individuals.

On the other hand, ethical climate encompasses a set of perceptions (beliefs and shared values) related to proper conduct and approaches to ethical issues (understanding and adhering to right and wrong). This climate shapes organizational decisions at all levels and under various conditions. The ethical climate not only significantly influences the behaviors of individuals and organizations but can also support and reinforce what most individuals inherently believe to be right. When the ethical environment is clear and positive, everyone knows how to act and respond during unavoidable ethical dilemmas, which are integral to work life (Khazeni et al., 2013). Bai and Vahedian (2023) found that an ethical environment plays a pivotal role in reducing nomophobia. According to them, this fact underscores the necessity for ethical leadership and robust ethical policies that can serve as a foundation for employee well-being in the digital era. Also, Hessari et al. (2022a; 2022b) point out the vital need to assess ways to alleviate the level of nomophobia in organizations.

Most of the research conducted on ethical climates is based on a theoretical framework developed by Victor and Cullen in the late 1980s to study the concept of ethical paths in profit organizations. Victor and Cullen define the ethical work climate as "common perceptions of organizational practices and procedures that have ethical content" (Victor & Cullen, 1988).

This climate comprises general organizational features that influence a wide range of organizational decisions, sometimes referred to as organizational culture. Victor and Cullen argue that individual characteristics alone are insufficient for describing or predicting ethical behavior. Their model consists of two dimensions, which, when combined, create various ethical climates and diverse ethical atmospheres.

The first dimension is related to Kohlberg's and Rest's theory regarding ethical criteria while also estimating organizational decision-making features. These criteria encompass self-interest, benevolence, and principled moral reasoning (Victor & Cullen, 1988). The second dimension supports the analysis utilized to generalize ethical criteria to individual, local (organizational, departmental, and workgroup), and global (external) levels (Stone & Henry, 2003). The individual analysis axis identifies the ethical reasoning sources among individuals. The local analysis identifies ethical reasoning sources within the organizational context. The global axis addresses ethical reasoning outside the organization (Maitland, 2005).

In the realm of ethical climates within organizations, the dynamics between personal interests, organizational interests, and professional values play a pivotal role in shaping the intricate balance between individual accountabilities and organizational objectives. Researchers such as Mahdavi and Ebrahimi (2015) have delved into this complex interplay, exploring how different ethical climates influence the conflict between the profession and the organization. Ethical climates that prioritize personal interests and align them with organizational goals are expected to yield a positive impact, ostensibly by reducing the conflict between the individual accountant's professional duties and the organization's demands. In these specific climates, there is a notable emphasis on advancing both personal and organizational interests. However, this orientation contradicts the fundamental norms and values within the accounting profession. In their professional capacity, accountants are guided by ethical principles that urge them to prioritize public interests above all else, reflecting a broader commitment to societal welfare.

Conversely, ethical climates that underscore social responsibility, adherence to laws, and strict compliance with professional codes of conduct are anticipated to influence the conflict between the profession and the organization negatively. These climates, grounded in ethical philosophies such as altruism and deontology, support the professional values upheld by accountants. Altruism emphasizes the maximization of common interests, aligning with the broader societal welfare, while deontology underscores the importance of strict adherence to laws and ethical codes. Therefore, when organizational climates align with these principles, they inadvertently challenge the divergence between individual professional values and organizational imperatives, potentially leading to conflicts.

To further contextualize these ethical climates, ethical criteria serve as the guiding principles underpinning organizational decision-making processes. These criteria are the moral compasses that organizations and individuals use to navigate ethical dilemmas. Ethical criteria can be broadly categorized into three groups: egoism, which revolves around maximizing individual interests; altruism, which focuses on the maximization of common interests and societal welfare; and deontology, which emphasizes adherence to laws and ethical codes. Moreover, the analysis dimension, a critical source of individual ethical reasoning, plays a significant role. This dimension can be categorized into three distinct scopes: individual, where considerations revolve around individual needs and preferences; local, where organizational interests take precedence; and global, where a broader perspective encompassing socio-economic system interests is considered (Smith et al., 2009).

In summary, the alignment between ethical climates, ethical criteria, and the analysis dimension within organizations intricately shapes the ethical landscape accountants navigate. Understanding these dynamics is essential for accountants and organizational leaders as they grapple with the challenge of harmonizing personal, professional, and organizational ethics within a cohesive framework. Recognizing the nuances of these ethical climates and criteria is crucial for fostering an ethical environment where accountants can fulfill their professional obligations while contributing meaningfully to the organizational objectives.

Ethical philosophy refers to the principles and rules considered during decision-making processes to distinguish between right and wrong actions. In other words, the theoretical focus of ethical climate is on ethical philosophy and socio-cognitive theory. The ethical criteria specified in the theory of ethical climate primarily include ethical theories, and they are inferred from Kohlberg's cognitive growth theory and, to a significant extent, from moral philosophy. Kohlberg identified three main types of ethical standards: egoism, altruism, and deontology. These standards also reflect the three primary categories of ethical theories: egoism, utilitarianism, and duty-based ethics.

According to the assertions of Victor and Cullen, the dimensions of ethical climate align with similar dimensions as Kohlberg's ethical standards and the three components of evolved moral theory. The framework proposed by Victor and Cullen refers to three criteria: egoism, benevolence, and laws. Research indicates that ultimately, among these three criteria, one prevails within an organization, becoming dominant and defining the organizational ethical climate (Filipoa, 2007).

Victor and Cullen (1988) propose that ethical climate structures are based on ethical philosophy. In their model, the "instrumental climate" corresponds to Kohlberg's "egoism" in moral philosophy. In the instrumental climate, employees solve ethical problems with a focus on their personal interests or organizational benefits. Victor's "caring climate" aligns with Kohlberg's "benevolence." In this climate, employees prioritize the well-being (profit and benefit) of others while solving ethical issues. The three ethical climates – "rules," "professionalism," and "independence" – are similar to Kohlberg's "laws." Victor and Cullen present various ethical climates within an organization, which include five dimensions: Interest, Professionalism, Rules, Independence, and Instrumentality (Tisayi, 2008).

The **Interest Ethical Climate** centers on utilitarian ethical criteria and is founded on genuine concern for the well-being of the organization's internal and external members who might be affected by their decisions. In this climate, individuals genuinely care about the goodness of their actions for the members inside and outside the organization.

The **Professionalism Ethical Climate** is associated with deontological ethical criteria. This dimension requires individuals to adhere to the rules and regulations of their profession. During decision-making situations, employees must base their decisions on the dictates of certain external systems such as laws, religious sources, and professional codes of conduct.

The **Rules of Ethics Climate** also aligns with deontological ethical criteria. This dimension pertains to the acceptance of legally prescribed behaviors by the institution. Organizational decisions are guided by a set of logical norms or standards, such as codes of conduct.

The **Independence Ethical Climate** encompasses deontological ethical criteria. In this dimension, individuals act based on their beliefs and values that a set of good principles has shaped.

The **Instrumentality Ethical Climate** includes egoism as an ethical criterion, primarily aiming to maximize personal interests. Individuals in this climate believe decisions made should serve the interests of organizations or personal gain. Various studies indicate that the Instrumentality climate has the least application in organizations compared to other ethical climates (Tisayi, 2008).

The pressure exerted on employees by superiors and peers stands as a significant challenge in the realm of organizational ethics. One striking revelation emerged from a study conducted in the United States, where managers were asked how much they felt pressured to sacrifice personal ethical standards to achieve organizational goals. Surprisingly, 64.4% of managers agreed with this statement, indicating a pervasive influence of pressure within organizational hierarchies. Delving deeper into the findings, a distinct pattern emerged across different management levels. Top managers exhibited a 50% agreement, middle managers demonstrated a 65% agreement, and operational managers were notably affected, with an 85% agreement rate. These numbers highlight a troubling trend: the lower the managers are in the hierarchy, the more pronounced the pressure becomes, suggesting a perceptual gap between top-level and lower-level managers regarding unethical behaviors.

This perceptual gap has profound implications, leading lower-level managers to engage in unethical practices due to various factors such as fear of retaliation, a misinterpretation of loyalty, or a distortion of job roles. The pressure to compromise ethical standards is evidently higher at operational levels, indicating a lack of comprehension or acknowledgment of the significant pressures faced by these managers by their superiors. Furthermore, the organizational culture is significantly shaped by the behaviors of superiors and peers. Instances of unethical actions and behaviors within management teams create a culture where employees evaluate the acceptability of their decisions based on their leaders' conduct.

Additionally, certain behaviors exhibited by superiors contribute to fostering an organizational culture that encourages unethical practices. Some managers adhere strictly to legal standards, believing that this alone fulfills their ethical duty. An operational mentality that prioritizes loyalty over ethical considerations can blur the lines of right and wrong. The lack of ethical leadership, where managers fail to take a leading role in doing what is right, adds to the complexity of the issue. Moreover, an overemphasis on profit-centric goals and evaluation systems can create an unsupportive environment, disregarding ethical considerations in pursuit of financial objectives. Insensitivity to the pressures exerted on subordinates and the absence of formal ethical policies exacerbate the problem, leading to a lack of management controls over ethical compliance and unclear guidelines on acceptable conduct.

In summary, the influence of superiors and peers on organizational ethics cannot be understated. The perceptual gap between different management levels, coupled with various behaviors exhibited by superiors, significantly impacts the ethical climate within organizations. Addressing these issues requires a holistic approach involving formal ethical policies and a cultural shift that promotes ethical leadership, transparency, and empathy within the organizational hierarchy. Managers must recognize the pressures faced by their subordinates and actively work towards fostering an environment where ethical decision-making is valued and supported at all levels of the organization. Bai et al. (2023) found that the ethical environment has a moderating role when it comes to investigating the effect of spiritual intelligence on job satisfaction, and this offers a nuanced perspective, pointing toward the significance of ethical practices in amplifying the positive effects of spiritual intelligence.

**Methodology and Theoretical Framework**

In this research, our approach was characterized by meticulous attention to detail and a commitment to scholarly rigor. We painstakingly constructed a robust conceptual framework, a visual representation of which can be found in Figure 1, through an exhaustive review of existing literature and an exploration of diverse theories. This conceptual framework not only served as the architectural blueprint of our study but also provided a guiding light, steering our research in the right direction. It formed the very bedrock upon which we built and tested our hypotheses, infusing our study with theoretical depth and methodological precision.

To uphold the ethical standards inherent in research, we crafted a formal letter outlining the specific objectives of our study. This letter was then judiciously sent to esteemed institutions specializing in business and entrepreneurship education in Tehran, seeking their approval and endorsement. A paramount concern throughout our research was the assurance of strict anonymity and confidentiality. We were unwavering in our commitment to safeguarding the individual data of our participants, recognizing the ethical imperative of this responsibility.

A significant challenge faced in research, the issue of non-response bias, was met with strategic and thoughtful countermeasures. We recognized the potential distortions that could arise from differences between survey participants and those who chose not to participate. To mitigate this bias, our methodology incorporated a series of targeted strategies. These included persistent follow-up reminders, enticing incentives, and a meticulous approach to sample selection. By deploying these tactics, we aimed to enhance response rates and minimize disparities, a methodology in alignment with the well-established principles advocated by Armstrong and Overton (1977).

Eight respected institutions' approval of our research objectives marked a crucial milestone. With their endorsement secured, our dedicated team engaged with a substantial sample of 185 faculty members across these institutions. During our interactions, we consistently emphasized the paramount importance of anonymity and confidentiality. This approach fostered trust and cooperation among the participants, resulting in the active engagement of 157 willing teachers. This extensive data collection effort spanned nearly two months, commencing on March 1st and concluding on April 20th, 2022. Remarkably, our endeavors yielded an impressive 85% response rate, with 140 meticulously completed questionnaires collected for in-depth analysis.

To ensure the robustness of our research, we employed rigorous analytical techniques. Confirmatory Factor Analysis (CFA) was utilized to evaluate the dimensions of our study scales meticulously. Additionally, descriptive statistics and correlation matrices were harnessed to validate and ensure the reliability of our dataset. Finally, Structural Equation Modelling (SEM) was applied, allowing us to scrutinize our hypotheses in intricate detail. Through these meticulous and thorough methods, we fortified the integrity of our study, ensuring the credibility and reliability of our findings.

Figure 1: Conceptual framework.

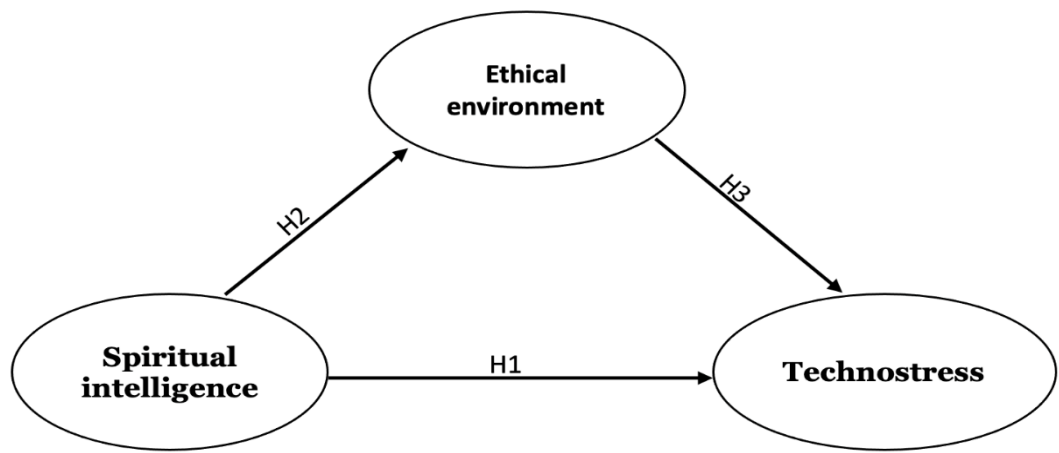

**Measures Employed in the Study**

During this research, a meticulously crafted set of questionnaires was deployed to extract invaluable insights, ensuring a thorough exploration of the research variables. To illuminate the ethical dimensions, a carefully curated 6-item questionnaire, rooted in Deshpande's Ethical Environment (1996), was expertly administered. This tool skilfully probed the ethical landscape, shedding light on the intricate interplay of values and principles within the study's context.

Furthermore, the study delved into the profound realm of spiritual intelligence, employing a robust 24-item scale questionnaire developed by King (2009). This nuanced instrument meticulously assessed the spiritual intelligence of the participants, offering profound insights into their inner beliefs, emotional connections, and overall spiritual well-being. Additionally, the Techno-Stress questionnaire by Tarafdar et al. (2007) was utilized, encompassing six questions on technology overload, three questions on technology invasion, five questions on technology complexity, five questions on technology insecurity, and four questions on technology uncertainty.

All responses, pivotal to the study's findings, were meticulously collected through a refined data-gathering process. Participants were invited to share their perspectives using a "7-point Likert scale ranging from 1= strongly disagree to 7 = strongly agree." This meticulous approach ensured the nuanced and accurate representation of participants' viewpoints, facilitating a comprehensive analysis of the collected data. Through the adept utilization of these diverse questionnaires and scales, this study was able to paint a detailed and multifaceted portrait of the participants' ethical, spiritual, and technostress experiences. This enriched the depth and validity of the research findings, offering a comprehensive understanding of the intricate interconnections between ethics, spirituality, and technology-related stress in the context of the study.

**Data Analysis**

**Measurement Model Analysis**
In our validation process, a meticulous and systematic approach was adopted, employing Confirmatory Factor Analyses (CFA) to assess the structure of our observed variables. To manage the high multicollinearity between constructs effectively, CFA was chosen, aligning with covariance-based structural equation modeling (CB-SEM) principles, as recommended by Hair et al. (2017). The analysis of our reflective constructs was executed using the lavaan package in R Studio, a method endorsed by Rosseel (2012). Additionally, specific aspects of our analysis were complemented by SPSS, ensuring a comprehensive and thorough evaluation of our data.

**Reliability and Validity Assessment**

**Reliability:** Our research instruments underwent a rigorous evaluation, demonstrating unwavering reliability. The internal consistency of the items was scrutinized using both Cronbach's alpha (CA) and composite reliability (CR). The values obtained, surpassing the 0.7 benchmark (refer to Table 2), signify the robustness of the measurement items. This meticulous examination ensures that items within each construct consistently measure the intended theoretical concepts, forming a strong foundation for subsequent analyses and interpretations.

**Convergent Validity:** Methodological rigor extended to assessing convergent validity. Through a thorough examination, all factor loadings were found to exceed the 0.70 threshold, confirming the constructs' ability to capture their underlying dimensions effectively. Furthermore, the Average Variance Extracted (AVE) values, surpassing 0.50 (as seen in Table 2), reinforced the convergent validity of the constructs. This rigorous validation confirms that measurement items converge to measure the same construct, ensuring the accuracy and consistency of research findings. These results underline the reliability and validity of the measurement model, affirming the integrity of the research methodology.

**Discriminant Validity:** A rigorous analysis was conducted to establish discriminant validity among latent variables. Variables with factor loadings exceeding 0.50 were meticulously identified and utilized to validate their distinctiveness from others (refer to Table 3). This scrutiny ensures that each variable stands apart from others, confirming their unique contribution to the research model. Attention to discriminant validity strengthens the study's robustness, confirming that the constructs under consideration are not only reliable and convergent but also distinct, thereby enhancing the overall quality and credibility of the research outcomes.

**Common Method Bias:** In behavioral studies, the influence of the measurement instrument on participants' responses, rather than their actual predispositions, can pose a threat known as Common Method Bias (CMB) (Podsakoff et al., 2003). Employing Harman's method, a widely recognized approach, we evaluated the factor score, which was remarkably low at 29.7%, well below the critical 50% threshold. This outcome indicates the minimal impact of CMB on our study, emphasizing the authenticity of participants' responses.

Table 2: Reliability.

| Constructs | CR | AVE | Cronbach's alpha | Confirmatory factor loadings range of items | Kurtosis (skew) |
|---|---|---|---|---|---|
| Spiritual intelligence | 0.83 | 0.75 | 0.9 | 0.825-0.900 | -0.19 (0.8) |
| Ethical environment | 0.73 | 0.67 | 0.84 | 0.790-0.875 | -1.02 (-0.30) |
| Technostress | 0.80 | 0.74 | 0.88 | 0.686-0.865 | -0.63 (-0.24) |

Table 3: Discriminant Validity.

|  | Spiritual intelligence | Ethical environment | Technostress |
|---|---|---|---|
| Spiritual intelligence |  |  |  |
| Ethical environment | 0.710 |  |  |
| Technostress | 0.423 | 0.395 |  |

**Analysis of the Measurement Model**

In our measurement model analysis, we employed the widely acknowledged method of confirmatory factor analysis (CFA) within the R Studio software, a tool respected in both social research and Information Systems (Kline, 2015). To ensure the credibility of our model, we adhered to stringent criteria. Our model displayed exceptional fit, surpassing established standards. Notably, our Comparative-Fit Index (CFI), Normed Fit Index (NFI), and Non-Normed Fit Index (NNFI) values exceeded 0.9, while Standardized Root Mean Square Residual (SRMR) and Root Mean Square Error of Approximation (RMSEA) values remained below 0.08, aligning precisely with the benchmarks defined by Hair (2019). Specifically, our RMSEA stood at 0.031, NFI at 0.91, CFI at 0.92, SRMR at 0.030, NNFI at 0.932, and TLI at 0.933, affirming the robustness and adequacy of our model.

**Results of Structural Equation Modelling**

In our structural equation modeling analysis, meticulously conducted using the lavaan package in R Studio, we integrated control variables to enhance the depth of our analysis. The evaluation of model fit involved essential indices, including the Comparative-Fit Index (CFI), Tucker-Lewis Index (TLI), Normed Fit Index (NFI), and Non-Normed Fit Index (NNFI). Our results yielded highly satisfactory values of 0.901, 0.883, 0.905, and 0.911, respectively. Additionally, our scrutiny included the Standardized Root Mean Square Residual (SRMR) and Root Mean Square Error of Approximation (RMSEA), both registering impressively low values of 0.03 and 0.031, adhering to the stringent criteria outlined by Hair et al. (2019). The Relative Chi-Square value, a vital indicator (chi-square/degrees of freedom), stood at 1.4, aligning impeccably with established standards for a good model fit, as advocated by Kline (2015). Importantly, our model not only met but exceeded these benchmarks, indicating an appropriate fit for our data. These validations are visually represented in Figure 2, clearly depicting the SEM outcomes derived from the R Studio software. This meticulous analysis underscores the reliability and robustness of our research findings.

Figure 2: Structural models with standardized estimates.

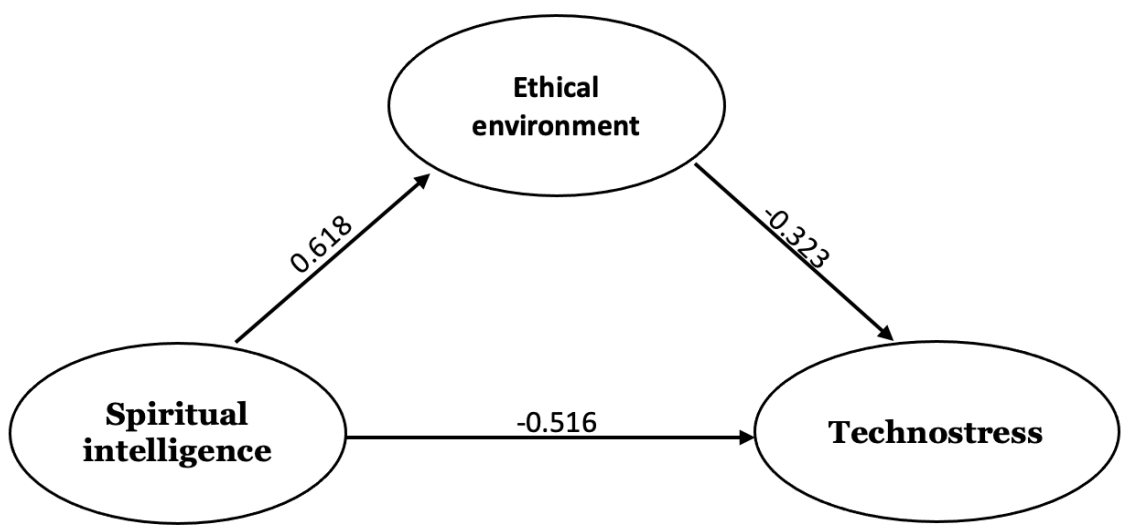

**Key Findings Overview**
In our comprehensive analysis, key findings emerged, validating our hypotheses and shedding light on essential connections within our research framework (refer to Table 4).

**Spiritual Intelligence and Technostress:** Our research unequivocally supports the first hypothesis. The analysis demonstrates a robust negative influence of spiritual intelligence on technostress ($p < 0.000$). This significant result highlights the profound positive impact of spiritual intelligence in alleviating employees' technostress, emphasizing its pivotal role in promoting psychological well-being amidst technological challenges.

**Ethical Environment and Technostress:** The second hypothesis was substantiated by delving into the relationship between the ethical environment and technostress ($p < 0.000$). This finding underscores the pivotal role of the ethical climate within an organization in mitigating employees' overall technostress. A nurturing ethical environment emerges as a powerful antidote to the adverse effects of technostress, fostering a healthier workplace atmosphere.

**Spiritual Intelligence and Ethical Environment Enhancement:** Furthermore, our research robustly supports the impact of spiritual intelligence on the ethical environment ($p < 0.000$). This result signifies a significant enhancement of the ethical climate within the organization due to heightened organizational commitment. With its positive influence, spiritual intelligence emerges as a catalyst for fostering a more ethical and supportive workplace milieu, further contributing to employees' well-being and job satisfaction. Bai et al., in their study on female employees, underscored the "profound influence of spiritual intelligence on job satisfaction." According to them, spiritual intelligence is not just a peripheral factor but central to understanding job satisfaction dynamics (2023).

Collectively, these findings underscore the intricate interplay between spiritual intelligence, the ethical environment, and technostress, offering valuable insights for organizational strategies aimed at promoting a harmonious, stress-free workplace environment.

Table 4: Hypothesis Testing.

| Path | Standardized coefficient | P-value | Result |
|---|---|---|---|
| Spiritual intelligence -> Technostress | -0.516 | 0.000 | Supported |
| Ethical Environment -> Technostress | -0.323 | 0.000 | Supported |
| Spiritual intelligence -> Ethical environment | 0.618 | 0.000 | Supported |

Mediating Role of Ethical Environment: Extending our investigation to the mediating effects, Table 5 illuminates the fourth hypothesis, confirming that the ethical environment indeed mediates the relationship between spiritual intelligence and technostress ($p < 0.000$). This nuanced understanding highlights the intricate interplay between spiritual intelligence, ethical environment, and technostress.

Table 5: Mediating Hypothesis Testing.

| Path | Mediating | Direct | Indirect | Total | Result |
|---|---|---|---|---|---|
| Spiritual intelligence -> technostress | Ethical Environment | -0.516 | -0.320 | -0.781 | Supported |

**Discussion**
The findings of this study highlight the significant negative impact of spiritual intelligence on technostress, underscoring the protective role that spiritual intelligence plays against the stress induced by technology. Spiritual intelligence, characterized by a deep understanding of personal meaning and values, appears to empower individuals to navigate the complexities and demands of technology-infused environments more effectively. This relationship is not only indicative of the buffer that spiritual intelligence can provide but also suggests that individuals with higher levels of spiritual intelligence are better equipped to integrate technology into their lives without the accompanying stress.

The role of the ethical environment in moderating the relationship between spiritual intelligence and technostress is particularly noteworthy. An ethical environment, marked by shared values, ethical leadership, and supportive policies, seems to provide a context in which the benefits of spiritual intelligence are amplified. In environments where ethical practices are emphasized, employees might find it easier to align their personal values with their work, reducing the conflict and stress that can arise from technological demands. This suggests that organizations investing in ethical climates may foster a more morally sound environment and help mitigate technostress's negative impacts.

**Practical Implications and Future Directions**

The findings of this study bear significant implications for organizational policy and practice. Given the pervasive nature of technology in the modern workplace, the insights regarding the mitigating effects of spiritual intelligence on technostress are particularly relevant. Organizations might consider strategies to cultivate spiritual intelligence among employees, such as mindfulness training, reflective practices, or workshops that emphasize personal values and meaning-making in professional contexts. These initiatives could help employees better integrate technology into their work lives, reducing the likelihood of technostress.

Furthermore, the importance of an ethical environment in mediating the relationship between spiritual intelligence and technostress points to the need for ethical leadership and robust ethical policies. Organizations should strive to create and maintain an ethical climate that supports employees' moral and spiritual values. This can be achieved through clear ethical guidelines, transparent decision-making processes, and leadership that exemplifies ethical behavior. Future research could explore the specific aspects of an ethical environment that are most effective in reducing technostress and enhancing the benefits of spiritual intelligence.

**Synthesis of Key Findings**

This study contributes to the growing body of literature on the interplay between spiritual intelligence, technostress, and the ethical environment within organizations. The findings offer a nuanced understanding of how spiritual intelligence serves as a buffer against the adverse effects of technology and how the ethical climate of the organization further influences this relationship. The study underscores the complexity of interactions between individual traits (like spiritual intelligence) and organizational factors (such as ethical environment) in shaping employees' experiences with technology in the workplace.

**Theoretical and Practical Contributions**

Theoretically, the study extends the discourse on spiritual intelligence by linking it to a contemporary challenge – technostress – in the organizational context. It also adds to the understanding of how organizational climates, specifically ethical environments, can modulate individual experiences of stress and well-being. Practically, the findings provide valuable insights for organizational leaders and HR practitioners. By highlighting the role of spiritual intelligence and ethical climates in mitigating technostress, the study suggests actionable strategies for organizations aiming to enhance employee well-being and productivity in a technology-dominated work environment.

**Future Research and Limitations**

While this study offers significant insights, it also has limitations that future research could address. The study's context, limited to specific educational institutions in Tehran, may affect the generalizability of the findings. Further research in diverse organizational settings and cultural contexts would enhance the understanding of these dynamics. Additionally, longitudinal studies could elucidate how the relationships between spiritual intelligence, technostress, and ethical environment evolve over time. Understanding these dynamics in greater depth could provide more comprehensive strategies for managing technostress in the digital age.